\documentstyle[11pt,newpasp,twoside,epsf]{article}
\def\edcomment#1{\iffalse\marginpar{\raggedright\sl#1\/}\else\relax\fi}
\reversemarginpar

\begin{document}
\title{Are Cluster Dwarfs Recycled Galaxies? }
 \author{Christopher J. Conselice\\ {\em California Institute of Technology, Mail Code 105-24 \\ Pasadena CA 91125; cc@astro.caltech.edu}}

\begin{abstract}
Although cluster dwarf galaxies are often neglected due to their faintness, 
recent observations demonstrate that they are critical for 
understanding the physical processes behind galaxy and cluster formation.  
Dwarfs are the 
most common galaxy type and are particularly abundant in clusters. Recent 
observational results suggest
that dwarfs in dense environments do not all form early in the universe,
as expected from hierarchical structure formation models.  Many of
these systems appear to be younger and more metal rich than dwarfs in 
lower density areas, suggesting they are possibly created by a tidal
process.  Several general galaxy cluster observations, including steep 
luminosity 
functions and the origin of intracluster light, are natural outcomes of
these processes.

\end{abstract}

\section{Introduction}

The most common galaxy in the universe are dwarf galaxies.  These
systems are typically faint (M$_{\rm B} > -18$), and have low
surface brightnesses ($\mu_{\rm B} > 23$ mag arcsec$^{2}$).  Since dwarfs are
so common, they are in every sense {\rm normal} galaxies.  The most
common type among dwarfs are dwarf ellipticals/spheroids, which
dominate the number density of galaxy clusters down to M$_{\rm B} = -11$
(Ferguson \& Bingelli 1994; Trentham et al. 2002).  Based on studies
of luminosity functions in clusters, there
are also more dwarf ellipticals per giant in denser regions than in the 
field.  This means, in the most simplest terms, that clusters of
galaxies cannot form through simple mergers 
of galaxy groups.  Some dwarf systems must form within the cluster
environment. The nature of this over-density may be the result of initial
conditions, or `non-standard' galaxy formation. That is, dwarfs 
may have formed after the cluster
was in place.  There is now ample evidence for this, the implications of
which can explain several galaxy cluster phenomenon.

The first clue that dwarfs are not produced through standard methodologies 
came from studies that showed dwarfs are generally found in abundance in 
dense areas such
as nearby clusters.   In fact, in standard CDM scenarios
dwarfs should be more common in lower density environments, but we see
the direct opposite (Trentham et al. 2002).  Dwarfs within dense
environments also have a broader distribution, both spatially and in terms of
their radial velocities, than giant ellipticals, similar to 
the pattern of infalling spirals.  This is an indication that both are  
recent additions to clusters (Conselice
 et al. 2001).  Internal dynamic evidence also suggests that at least some
dwarfs are rotating (e.g., Geha et al. 2003), a feature not seen in Local
Group dEs.  Finally, the stellar populations of some faint M$_{\rm B} > -15$ 
dwarf 
galaxies appear to be metal rich (near solar), implying that the dwarf
population itself is inhomogeneous and may have multiple origins.  We discuss
several scenarios which have been proposed to explain these observations and
argue that one method which can reproduce these trends involves
the formation of dwarfs from already existing galaxies through a tidal effect.

\section{Dwarf Galaxy Properties}

The two places where dwarfs have been studied in detail are in nearby
rich clusters, such as Virgo, Fornax and Coma, and in the Local Group.  Most
of the data we discuss therefore comes from these sources.  In particular
a significant fraction of the data presented in this article comes from the 
papers by
Conselice, Gallagher \& Wyse (2001,2002,2003) and Conselice et al. (2003b).
We list below our current understanding of dwarf properties in 
terms of various observational quantities.

\noindent {\bf Number Densities}:  The
luminosity function (LF) in all galaxy environments are dominated
by dwarf galaxies. The over-density of dwarfs in clusters
is about 5-10 times that in groups.  Another way to quantify this is through
the faint end slope of the LF, $\alpha$.  The value
of $\alpha$ in rich clusters, such as Virgo is around $\alpha = -1.4$ (Sandage
et al. 1985; Conselice et al. 2002), with
some results suggesting even steeper LFs with
$\alpha = -1.6$ (Sabatini et al. 2003).   This is steeper than field
values, such as in the Local Group ($\alpha = -1.1$; 
van den Bergh 2000), yet flatter than the value predicted by CDM for all
environments ($\alpha = -2$).  Environment
clearly affects the way these systems are produced, which is generally
not a prediction in CDM models (cf. Tully et al. 2002).

\noindent {\bf Spatial Positions}: While Local Group dwarf galaxies, particularly
dwarf ellipticals, are strongly clustered around giant galaxies
in the Local Group (e.g., van den Bergh 2000), the opposite is found for 
low-mass galaxies in clusters, where most are neither clustered around, nor 
distributed globally similar to,  giant elliptical galaxies 
(Conselice et al. 2001).  Both dwarf ellipticals and irregulars also
have a broader distribution in clusters, that is they are not clustered
towards the center, but are spread throughout (e.g., Conselice et al. 2001).  


\begin{figure}
\vskip -2cm
\plotfiddle{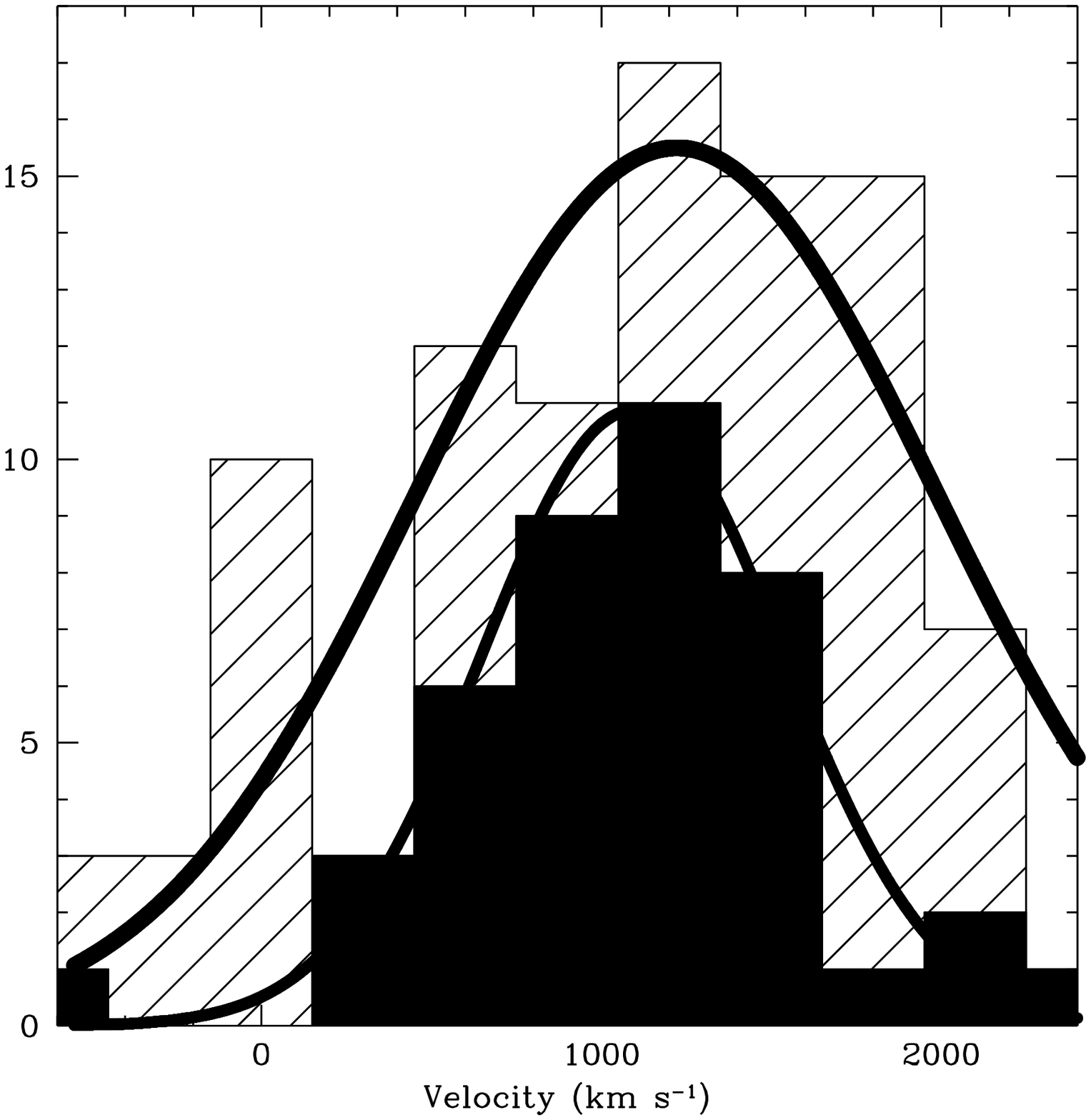}{0.5cm}{0}{30}{30}{-190}{-250}
\plotfiddle{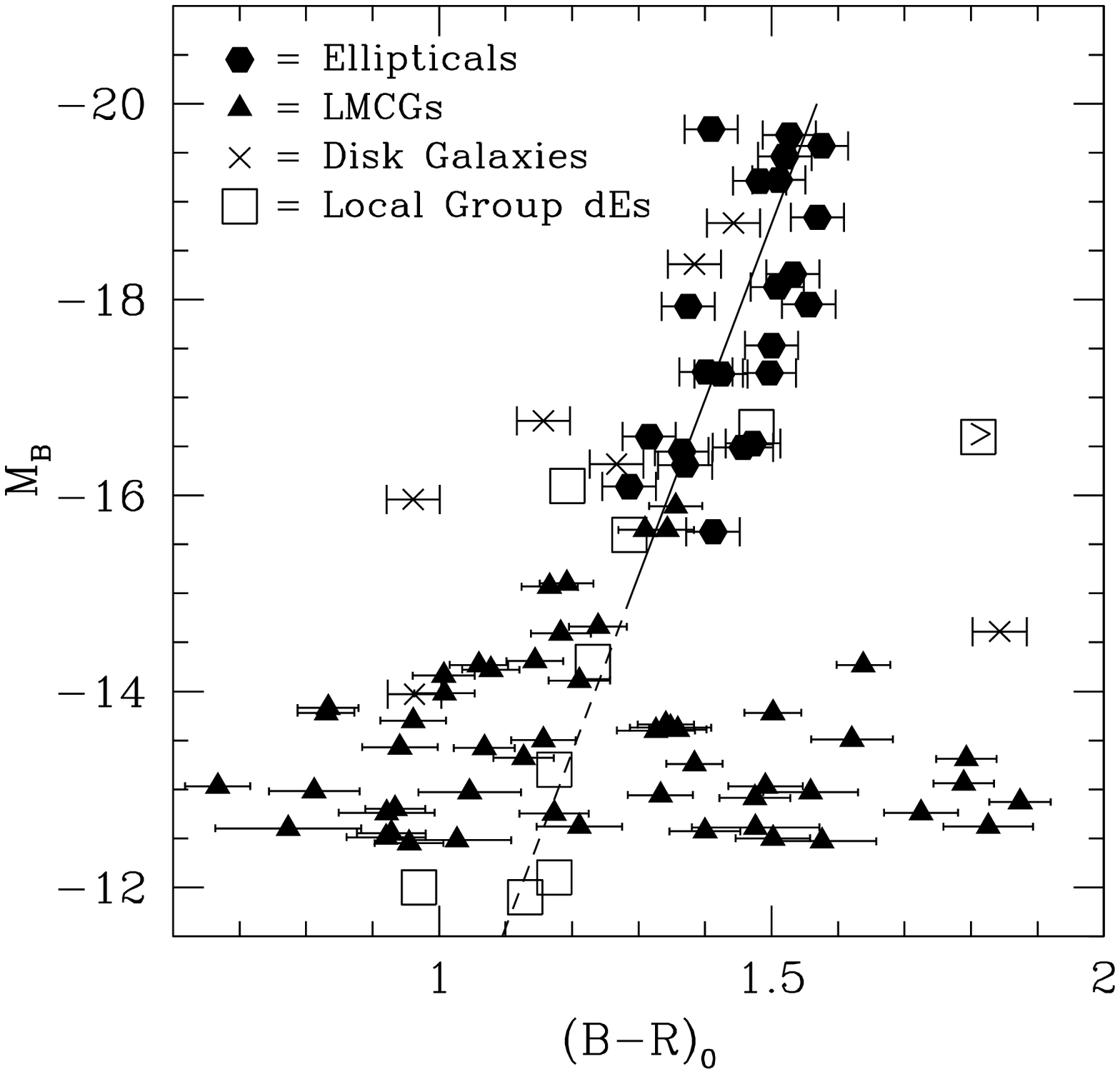}{0.5cm}{0}{34}{34}{0}{-235}
\vskip 6.5cm
\caption{(a) Velocity histograms for giant ellipticals (solid) and dwarf 
ellipticals (shaded) in the Virgo cluster (Conselice et al. 2001). (b) 
Color magnitude diagram for galaxies in the Perseus cluster, 
demonstrating the large color scatter for systems with M$_{\rm B} > -15$.
The solid boxes are where Local Group dEs/dSphs would fit on this plot. Dwarf
ellipticals are labeled as low mass cluster galaxies (LMCGs).}
\end{figure}

\noindent{\bf Radial Velocities}: The radial velocities of low-mass cluster 
galaxies, including S0s, spirals, dwarf irregulars and dwarf
ellipticals have a wider distribution than the ellipticals (see
Figure~1a).  For example, Virgo cluster elliptical
galaxies have a narrow Gaussian velocity distribution, with $\sigma = 462$ 
km s$^{-1}$, concentrated at the mean 
radial velocity of the cluster.  The other populations, including
the over 100 classified dwarf ellipticals in Virgo with radial
velocities, have much broader, and non-Gaussian, velocity distributions 
($\sigma \sim 700$ km s$^{-1}$), all with velocity dispersion ratios with 
the ellipticals consistent with their being accreted (e.g., 
Conselice et al. 2001).  There is also significant sub-structure within
these velocity distributions, unlike the case for the giant
ellipticals.

\noindent{\bf Stellar Populations}:   Currently,
we know with some certainty that dwarf galaxies have either 
young/metal rich or old/metal poor stellar populations 
(e.g.,  Poggianti et al. 2001).  This is further seen
in complete color-magnitude diagrams in nearby rich clusters,
such as Perseus,  down to M$_{\rm B} = -12$ (Conselice et al. 2002, 
2003a).  Fainter dwarfs also tend to be even more heterogeneous, with a large
scatter from the color-magnitude relationship (CMR) (Conselice et al.
2003a; Rakos et al. 2001) (Figure~1b).  This trend is found in
several nearby clusters, including Fornax, Coma and Perseus, and can be 
explained by different dwarfs having mixtures of ages and 
metallicities (e.g., Poggianti et al. 2001; Rakos et al. 2001; 
Conselice et al. 2003a).  Stromgren
and broad-band photometry reveal that the redder dwarfs are metal enriched
systems (Figure~2).   

\noindent {\bf Internal Kinematics}:    One key observational result,
predicted in the simulations of Moore et al. (1998), is that a dwarf
elliptical which has been transformed from a spiral should reveal some
rotation.  Using 8-10 meter class telescopes, the evidence is ambiguous
with some dEs showing rotation (Pedraz et al. 2002), while others
clearly do not (Geha et al. 2003).  There is no obvious difference between
these two populations in terms of morphology or chemical abundances 
(Geha et al. 2003), although these samples are still very small.  The
central velocity dispersions of these systems is also quite low, indicating
that dark matter does not dominate, at least in their centers.

\begin{figure}
\vskip -2.8cm
\plotfiddle{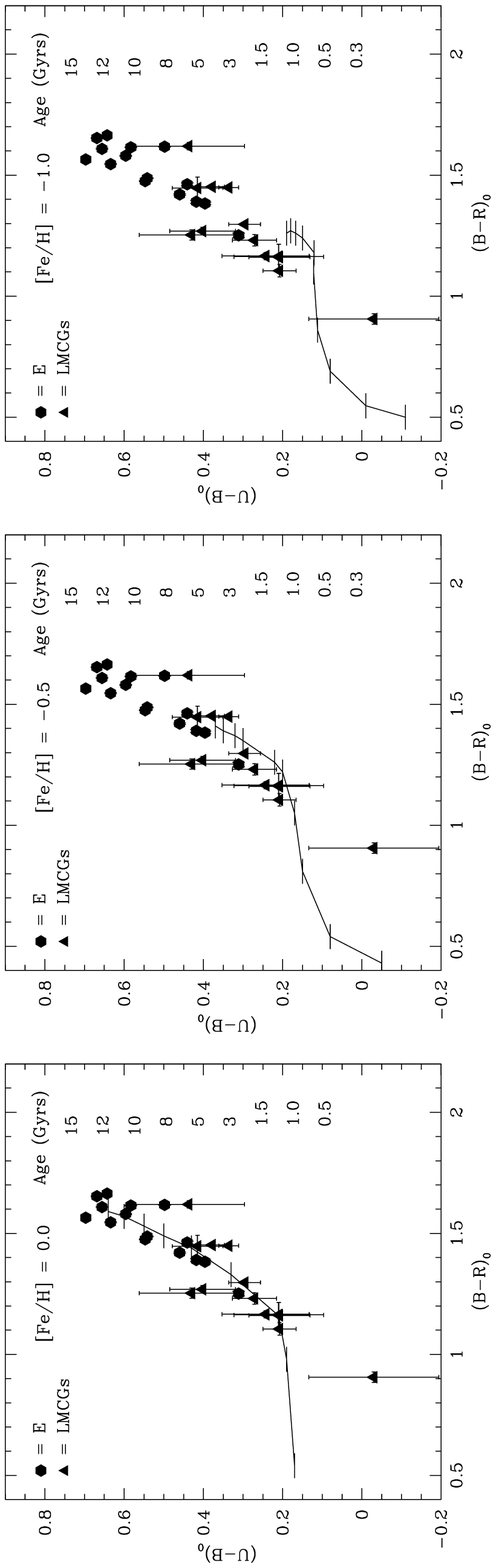}{0.5cm}{-90}{50}{50}{-200}{30}
\vskip 6.5cm
\caption{Three stellar synthesis modeled age tracks on a UBR color-color diagram 
at constant metallicities of solar, [Fe/H] = -0.5, and [Fe/H] = -1. The age
range is 0.3 Gyrs to 15 Gyrs for the [Fe/H] = -0.5 and -1
models and 0.5 Gyrs to 15 Gyrs for the solar metallicity models. Dwarf
ellipticals are labeled as low mass cluster galaxies (LMCGs). }
\end{figure}

\section{Dwarf Galaxy Origins}

Any successful theory of dwarf galaxies, particular 
dwarf ellipticals, must therefore explain the following properties: over-dense in
relation to massive galaxies in rich environments, mix of stellar populations,
ability to survive in dense environments, diffuse spatial and velocity
structure, and mixtures of rotation.  This theory must also explain why fainter
dwarfs are more heterogeneous than brighter ones.

One idea recently suggested by Tully et al. (2002) is that the dark matter
halos of dwarfs are `squelched' in lower density environments due to a large
ultraviolet background after the universe was reionized.  This explains the
differences in $\alpha$ between different environments, but does not explain
how within the same environment there is a great diversity in the dwarf
population. It also does not easily explain why many dwarfs appear to have
recently been accreted into clusters.   In simple collapse + 
feedback scenarios (Dekel \& Silk 1986), dwarfs are formed when gas collapses 
and forms stars. These stars produce winds that expels gas from these systems,
halting any future star formation.
In this formation scenario dwarfs formed before the cluster
ellipticals, or at least formed within groups that later merged to form 
clusters. Fainter dwarfs 
however, cannot all be born in groups which were later accreted into 
clusters along with the massive galaxies, due to the high dwarf to giant 
galaxy ratio found in clusters (Conselice et al. 2001, 2003a).
The above evidence suggests that simple low-mass galaxy formation scenarios
can be safely ruled out for some dwarfs.
One alternative idea is that some modern dwarfs
formed after the cluster itself was in place by collapsing out
of enriched intracluster gas.  Another is that the intracluster medium (ICM) 
is able to retain enriched gas that in the Dekel and Silk
(1986) paradigm would be ejected by feedback, but remains due to the 
confinement pressure of the ICM (Babul \& Rees 1992).  This scenario would 
explain the higher metallicities of some of the fainter dwarfs.

An alternative scenario, now gaining in popularity, is that dwarfs form in 
clusters through
a tidal origin.   Two main possibilities for this are tidal dwarfs (Duc
\& Mirabel 1994), and as the remnants of 
stripped disks or dwarf irregulars (Conselice et al. 2003a).
The velocity and spatial distributions of dwarfs suggests that they were likely
accreted into clusters during the last few Gyrs 
(Conselice et al. 2001).  This, combined
with the high metallicities of these cluster dwarfs, and the fact that their
stellar populations are fundamentally different than field dwarfs (e.g.,
Conselice et al. 2003a; Figure 1b) suggests that the cluster environment has
morphologically transformed, or stripped, accreted galaxy material into dwarfs.
There is evidence for this process currently occurring in nearby clusters 
(e.g., Conselice \& Gallagher 1999).  
If dEs form from infalling spirals then there should also
be systems now being transformed which appear morphologically as dEs, but
retain some of the gas left over from their precursor.  These systems
would only exist in the outer parts of clusters, as any that venture towards the
core will be rapidly stripped of material.  Very deep Arecibo observations
of Virgo dEs reveal that $\sim 15$\% of a sample of 56 Virgo dEs have
HI detections, all of which are located outside the core of
the cluster (Conselice et al. 2003b).  Other detailed morphological 
investigations of nearby cluster
dwarf ellipticals show that they contain a wide diversity of structures,
some with tidal features, and others with apparent spiral structures (e.g., 
De Rijcke et al. 2003; Barazza et al. 2003; Graham et al. 2003).

\section{Implications of Dwarf Recycling}

Environment likely plays a role in all aspects of galaxy formation and
evolution. If indeed these objects were formed after the cluster, due to
a tidal process, there are several features of clusters that this can possibly
explain.  The first is that the LF of clusters will change
after these lower mass galaxies form.   The LF of
galaxies in the Perseus cluster becomes flatter, with a similar
faint end slope ($\alpha$) as the field, after removing these red galaxies 
(Figure~3).

\begin{figure}
\vskip -2.8cm
\plotfiddle{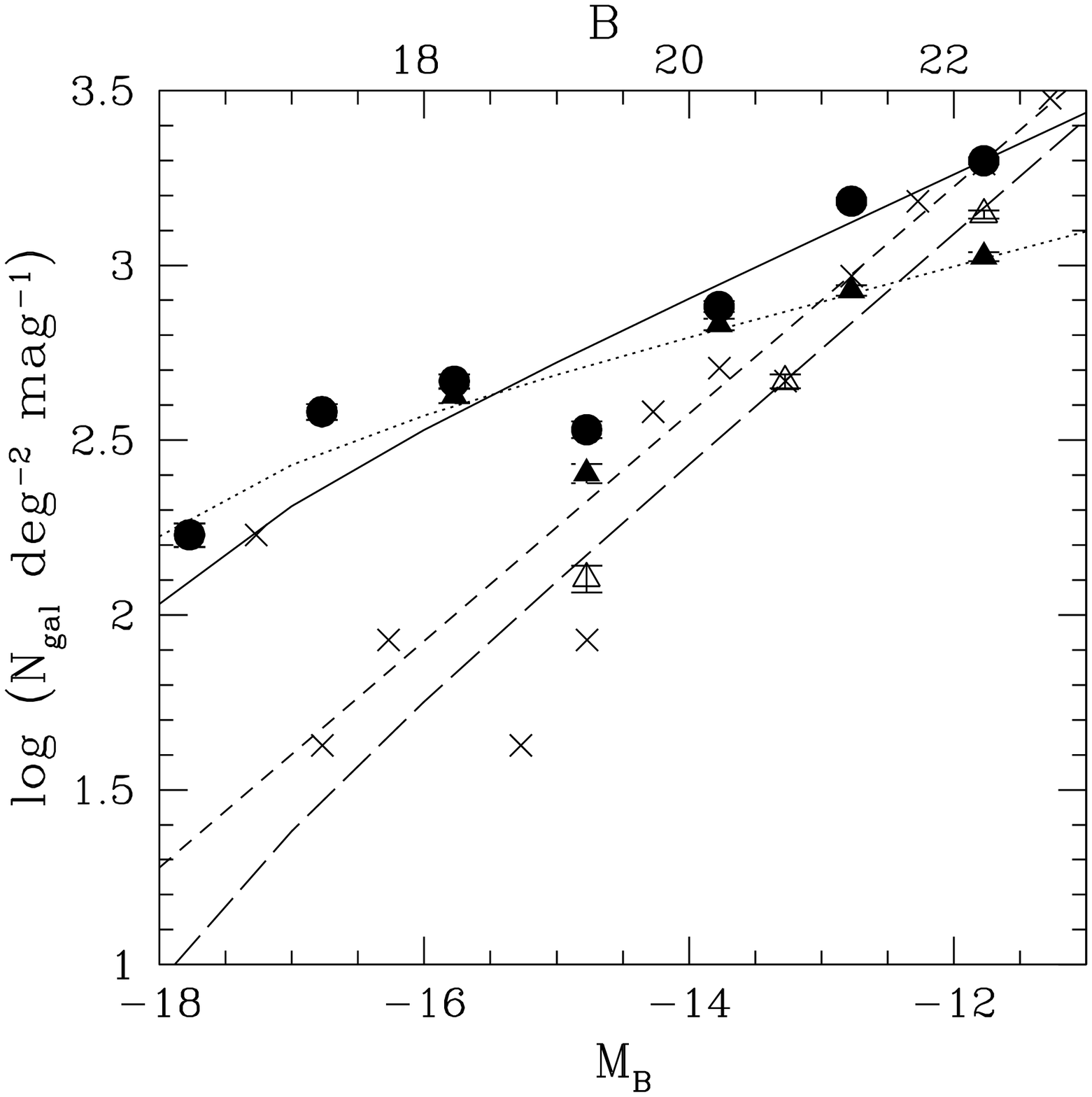}{0.5cm}{0}{30}{30}{-190}{-250}
\plotfiddle{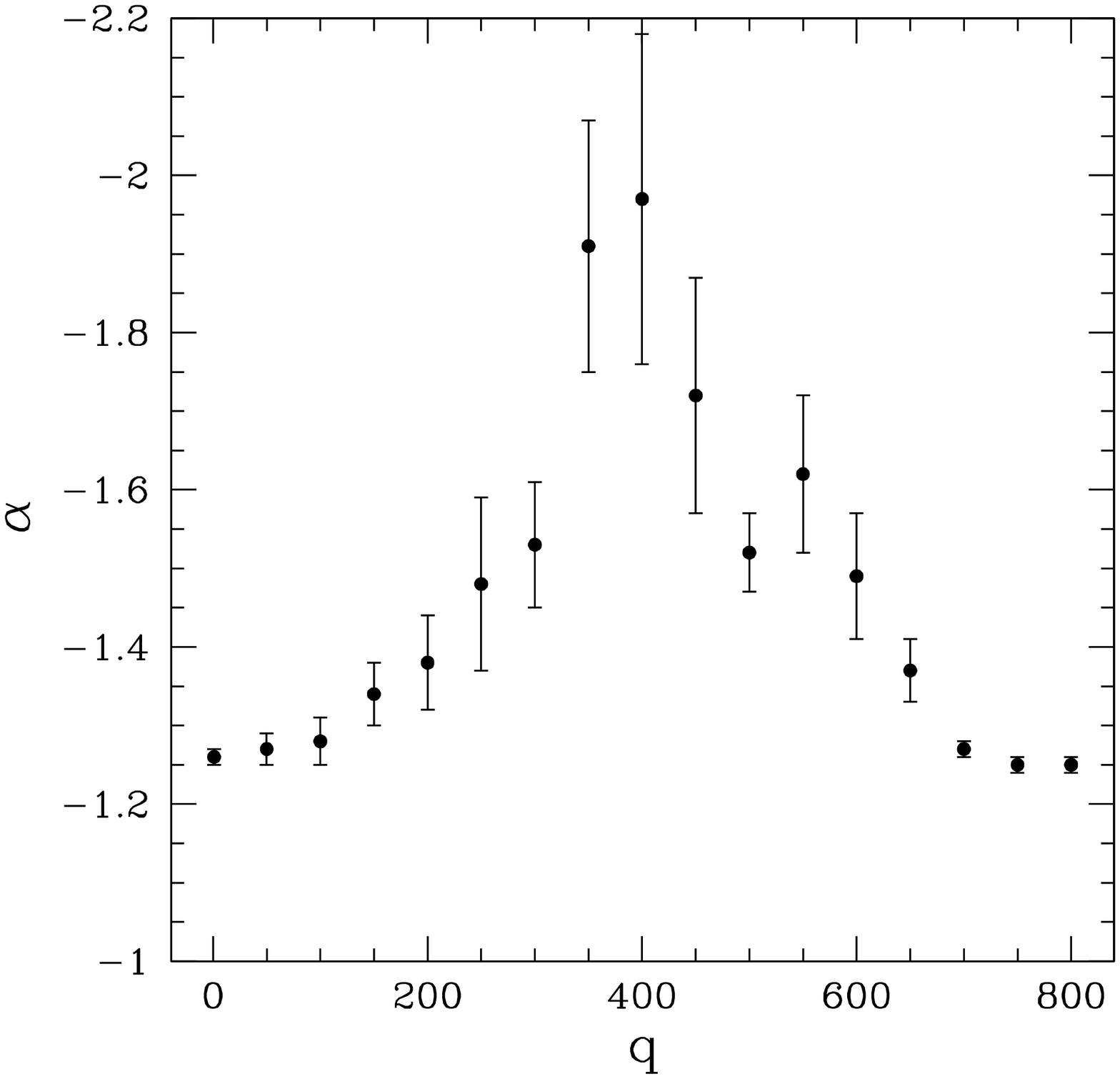}{0.5cm}{0}{30}{30}{0}{-230}
\vskip 6.5cm
\caption{(a) The Perseus cluster luminosity function plotted in
various ways down to $M_{B} = -11$.  The solid round points, 
and fitted solid line, is the total luminosity function of the central region 
of Perseus.  The luminosity function for dEs redder and bluer than the CMR
prediction are plotted as open and solid triangles and long dashed and
dotted lines.  The crosses mark the density of background galaxies. (b)
Modeled cluster luminosity function
slope, $\alpha$, as a function of the number of high-speed maximum 
interactions 
(q $\sim$ time) cluster galaxies undergo during evolution in a Perseus like 
cluster (see Conselice 2002).}
\end{figure}

If these systems originate from tidally disturbed galaxies then
the amount of light liberated is enough to account for all intracluster
light (Conselice et al. 2003a).  For example, the total luminosity of 
intracluster light in the Virgo Cluster within 2$^{\circ}\,$ of M87 is 
2 $\times 10^{11}$ L$_{\odot}$. If this material originates from tidally 
striped 
objects whose remnants are dwarfs, we can compute how much material on average 
each must have lost.  There are 170 dEs, 148 dE,Ns, and 14 S0 galaxies within 
this radius in Virgo. On average, if all the dE and dE,N galaxies are remnants 
of stripped galaxies then $0.6 \times 10^{9}$ L$_{\odot}$ was lost by each 
object. 
If we consider only dE galaxies as remnants of this process then $1.1 \times 
10^{9}$ L$_{\odot}$ was lost by each dE. This is enough material to suggest that 
each dwarf might have been a moderate-sized disk galaxy in the past.

\acknowledgments{I acknowledge my collaborators: Jay Gallagher, Rosie Wyse and
Karen O'Neil for their participation in the projects described here. I also
thank the organizers of this symposium for creating an interesting meeting, and
for their patience and support. }

\end{document}